\documentclass{aa}

\usepackage{graphicx}
\usepackage{natbib}
\bibpunct{(}{)}{;}{a}{}{,}
\usepackage{times}

\newcommand{\reduceme}{\mbox{R\raisebox{-0.35ex}{E}D%
\hspace{-0.05em}\raisebox{0.85ex}{uc}\hspace{-0.90em}%
\raisebox{-.35ex}{{m}}\hspace{0.05em}E}}

\newcommand{\OIII}{[O\,{\sc iii}]}

\begin{document}

\title{Stellar Population Gradients in Bulges along the Hubble Sequence. \thanks{Based on observations collected at the Roque de los
  Muchachos Observatory and at the European Southern Observatory, proposal
  numbers 58.A-0192(A), 59.A-0774(A) and 61.A-0326(A)}}

\subtitle{I. The Data}

\author{J. Gorgas\inst{1}      \and     P. Jablonka\inst{2}
    \and      P. Goudfrooij\inst{3} }

%   \offprints{J. Gorgas}

\institute{Dpto. de Astrof\'{\i}sica, Facultad de F\'{\i}sicas,
 Universidad Complutense de Madrid, E-28040, Madrid, Spain
% \email{fjg@astrax.fis.ucm.es}
\and
Observatoire de l'Universit\'e de Gen\`eve, Laboratoire d'Astrophysique de
  l'Ecole Polytechnique F\'ed\'erale de Lausanne (EPFL), CH-1290 Sauverny,
  Switzerland
\and
 Space Telescope Science Institute, 3700 San Martin Drive, Baltimore,
 MD 21218, USA
}

\date{}  \abstract{ This is the  first  paper presenting our long-term
  project aimed at studying the nature  of bulges through the analysis
  of   their stellar   population  gradients.  We   present  deep
  spectroscopic  observations   along  the minor   axis  and  the data
  reduction for a  sample of 32 bulges  of edge-on spiral galaxies. We
  explain    in detail our    procedures  to  measure their  dynamical
  parameters (rotation  curves and  velocity dispersion  profiles) and
  line-strength indices,  including  the  conversion to   the Lick/IDS
  system.  Tables giving the  values of  the  dynamical parameters and
  line-strength indices at  each  galactocentric radius  are presented
  (in electronic  form) for  each  galaxy of the  sample.  The derived
  line-strength gradients   from this dataset   will be analyzed  in a
  forthcoming paper to set  constraints on the different scenarios for
  the formation of the bulges.

\keywords{Galaxies: bulges  -- Galaxies: stellar content -- Methods:
  data analysis }}

\maketitle

\section{Introduction}

Bulges of spiral galaxies are cornerstones for constraining theories
of galaxy formation. Located at the centres of spiral galaxies, they
hold the signature of the sequence of formation of the different
sub-systems building a spiral galaxy:\ halo, disc, and bulge. While
the mass of disks of spiral galaxies in the local universe is
approximately constant among spirals of different Hubble types (e.g.,
Arimoto \& Jablonka 1991), the prominence of bulges varies widely
along the Hubble sequence. Hence, bulges are likely to hold important
clues to our understanding of galaxy evolution and the nature of the
Hubble sequence.

Formation scenarios for bulges can be divided into two categories.
One of them states that bulges are formed in a way similar to that
of low- to  intermediate-mass  elliptical    galaxies,  based on
the    strong similarities  between    global  properties of many
bulges   and of elliptical galaxies. Bulges  and (small)
ellipticals populate the same location in the  Fundamental Plane
\citep*{bend+92,falc+02},  and they form  a continuous sequence in
the $V_{max}/\sigma$ vs.\ ellipticity diagram  \citep[being
supported by rotation;][]{bend+92}. Furthermore, spectroscopic
studies  of the central part  of bulges have  shown that their
mass-metallicity relation  (when derived from $\alpha$ elements,
e.g.\ the Mg$_2$ index) is consistent with that of elliptical
galaxies \citep*{jabl+96,idia+96,moohol06}, but see also Prugniel,
Maubon, \& Simien (2001), and Proctor \& Sansom (2002). We refer
the reader to the introduction in Jablonka, Gorgas, \& Goudfrooij
(2007, hereafter Paper~II) for a discussion on previous
spectroscopic studies of bulges.

The other popular type of formation scenario for bulges is the
``secular evolution'' scenario in which bulges form from disk
material through redistribution of angular momentum. This scenario
involves stellar bars driving gas from the disk into the central
region of the galaxy, thus triggering star formation
\citep*[e.g.,][]{pfenor90,norm+96,korken04}. If enough mass is
accreted,  the bar itself will    dissolve  and the  orbits of   stars
involved    in the  process     will   yield  a bulge-like     spatial
distribution.  Galaxies would thus evolve  from  late to earlier types
along the Hubble sequence. Observational support for secular evolution
has been provided mainly for  late-type spiral galaxies ($\sim$ Sb and
later), including correlations between the scale lengths of bulges and
disks \citep*{cour+96,mcar+03} and  the  radial light  distribution of
bulges  of   late-type spirals  being closer   to exponential  than to
classical de Vaucouleurs' $R^{1/4}$ profiles \citep*{balc+03,dejo+04}.

Stellar    population   studies should be   able    to place important
constraints on  the  formation mechanisms  for bulges.  If bulges form
mainly  through  dissipative  collapse  without significant subsequent
merging, simulations indicate  that  the slope of   radial metallicity
gradients should  steepen with increasing   galaxy mass and luminosity
\citep{carl84}.  Simulations of dissipative  merging of gas-rich disks
has also been shown to produce metallicity gradients that steepen with
increasing mass \citep[e.g.,][]{bekshi99}.   Conversely, the impact of
secular evolution  on population gradients   is expected to be  rather
different. Simulations including   gas and effects  of  star formation
show that disk gas fueled  to the central   regions renders a  central
region that  is  younger and more metal  rich  than the outer regions,
whereas metallicity  gradients outside the  central region flatten out
significantly $\sim$\,1 Gyr after formation of a bar, both for gas and
stars \citep*{frie+94,frie98}.  Observational evidence for this effect
exists,  albeit so  far  only for the  gas  component in spiral disks.
Several studies have  shown  that global radial  gradients of  the gas
metallicity in barred spirals are shallower than gradients in unbarred
spirals of the same Hubble type \citep*{viledm92,zari+94,marroy94}.

In  terms  of {\it stellar\/}  population   gradients in  bulges,
past studies  mainly  used broad-band  imaging.   The studies of
\citet{balpel94} and  \citet{pele+99} used  optical and
near-infrared colors and  found that color gradients of  luminous
bulges  increased with  luminosity whereas  faint bulges showed
stronger gradients than  expected  from the trend observed for the
luminous   bulges.  However, information  in  colors  is generally
rather degenerate in age, metallicity, and/or extinction.

Spectroscopic studies allow the  measurement of line  strength
indices that  are   insensitive  to dust  extinction,   and  allow
a cleaner separation   of   age and   metallicity    of  a stellar
population. While line   strength gradients have  been measured
for many elliptical and S0 galaxies using the Lick/IDS system of
indices
\citep*[e.g.,][]{caro+93,gongor96,fish+96,mehl+03,kunt+06}, such
data are relatively hard to acquire for bulges of spirals at high
enough signal-to-noise ratios,  given the often low surface
brightness of bulges and  the difficulty  of  disentangling the
observed  spectra into bulge  and  disk  components. This   is
especially difficult  for late-type spirals in face-on or
moderately inclined configurations.

With this in mind, we embarked on an extensive spectroscopic
survey of 32 spiral galaxies in  an edge-on configuration to avoid
contamination by  disk light, and  with Hubble  types ranging from
S0 to  Sc. Early results from subsets  of  the data acquired
during this survey  were published  in \citet{goud+99},
\citet{jabl+02}, and \citet{Gorg+03}. In this first paper  on the
final and  comprehensive results of  this survey,  we present the
extensive  data  reduction   and  analysis procedures leading to
the  final  line strength measurements.     The galaxy sample is
presented in  Section~\ref{s:sample}. Section~\ref{s:obs}
describes the observations, and Section~\ref{s:reduction} provides
details on the data reduction procedures. Section~\ref{s:dyn}
contains a description  of   the determination of dynamical
parameters (radial velocities and velocity dispersions   as a
function of bulge   radius), and the line strength index
measurements themselves       are described     in
Section~\ref{s:indices}. Finally, Section~\ref{s:summ} contains  a
summary of this work.

%%%%%%%%%%%%%%%%%%%%%%%%%%%%%%%%%%%%%%%%%%%%%%%%%%%%%%%%%%%%%%%%%%%%%%%%%
\section{The sample} \label{s:sample}

\begin{figure}
\resizebox{1.0\hsize}{!}{\includegraphics[angle=-90]{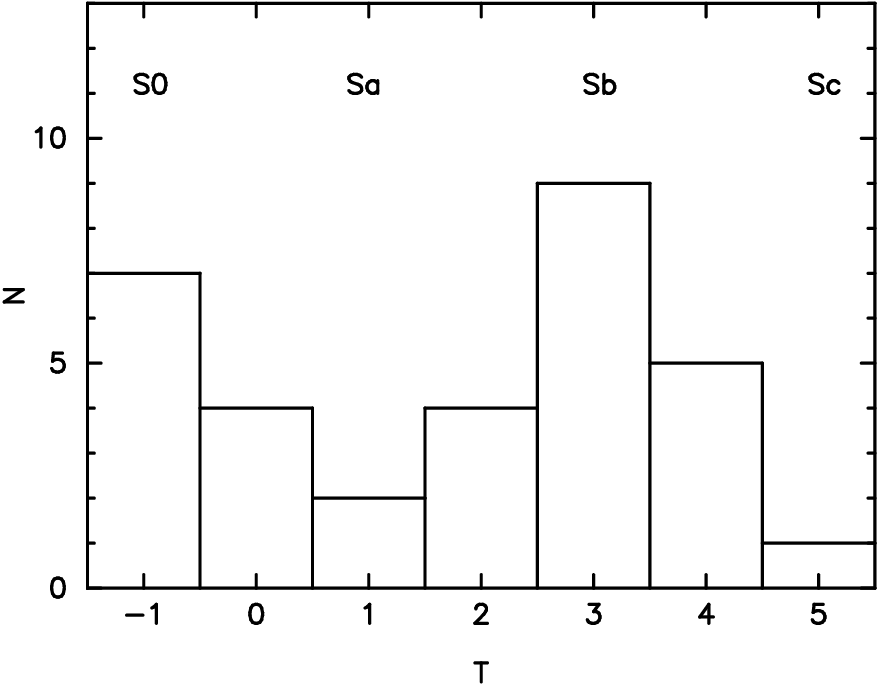}}
 \caption{Histogram showing the distribution in Hubble types of the galaxy sample.}
 \label{phisto}
\end{figure}

We  selected a  sample of  32  genuine (or  close   to) edge-on spiral
galaxies. Galaxies in the  northern hemisphere were selected  from the
Uppsala General Catalogue (Nilson 1973),  while southern galaxies were
selected  from the ESO/Uppsala catalogue (Lauberts  1982). As shown in
Figure~\ref{phisto}, the  galaxies span a large  range in Hubble types
(from  S0  to Sc).  Given  the  edge-on configuration of the galaxies,
precise morphological  classification is  difficult.  Due  to  this we
consider a Hubble T-type uncertainty of $\pm$ 1,

which   is also  a fair representation   of the catalogue-to-catalogue
variations for a given galaxy.

All the  galaxies are nearby objects,  with radial  velocities
between 550  and 6200 km~s$^{-1}$, and  were chosen to  cover a
range in bulge luminosity ($-17.5>M_{\rm V}>-20.5$).
Table~\ref{tgalaxies} lists the galaxy sample and some relevant
parameters, including the corresponding bulge types (from
L\"utticke et al.\  2000) and the local environment, which also
varies  significantly among the sample   galaxies. A few galaxies
have previous photometric and   spectroscopic observations from
other authors (see Section~\ref{s:indices}) and were included for
comparison purposes.

\begin{table*}[t]
\caption{The galaxy sample.} \label{tgalaxies}
 \centering
\begin{tabular}{llrrccccrrrrr} \hline \hline
 & Type  &  \multicolumn{1}{c}{$T$}  & \multicolumn{1}{c}{$B_{\rm T}$} &  $i$
& Bulge Type$^{\rm a}$ & Environment & run$^{\rm b}$ &  PA$^{\rm
c}$ & \multicolumn{1}{c}{exp.$^{\rm d}$} &
\multicolumn{1}{c}{$r_{\rm eff}^{\rm e}$} &
\multicolumn{1}{c}{$\Delta r_{\rm eff}$} &
\multicolumn{1}{c}{$V_{\rm max}^{\rm f}$} \\
\hline
NGC  522     &  Sc &   4.1 &    13.98   &  90& box         &             &2    &  123 &   8200 &3.5&0.7&196.1\\
NGC  585     &  Sa  &   1.0 &    14.06   &  90& close to box&             &4,5  &   86 &   7200 &2.4&0.7&246.6\\
NGC  678     &  SBb &   3.0 &    13.31   &  90& box         & cluster     &1    &  168 &  12000 &23.3&1.4&169.0\\
NGC  891     &  SAb  &   3.0 &    10.83   &  88& box         &             &1    &  112 &  10500 &13.2&0.7&212.0\\
NGC  973     &  Sbc  &   3.2 &    14.17   &  90& box         &             &2    &  138 &  13800 &3.4&0.4&270.0\\
NGC 1032     &  S0-a&   0.4 &    12.74   &  90& elliptical  & companions? &1    &  158 &   9000 &11.4&1.1&283.9\\
NGC 1184     &  S0-a&   0.0 &    12.95   &  90& close to box&             &2    &   79 &   8200 &11.3&1.9&\\
NGC 1351A    &  SBbc&   4.5 &    14.16   &  80& elliptical   & cluster     &5    &  132 &   5400 &5.8&0.8&90.7\\
NGC 1886     &  Sab &   3.9 &    13.80   &  90& peanut      & isolated    &3    &   60 &  10800 &1.6&0.3&154.6\\
NGC 3957     &  SA0 &  $-1.0$ &  13.03   &  90& close to box& cluster     &3    &  173 &   7200 &4.0&0.8& \\
NGC 5084     &  S0  &  $-1.8$ &  11.55   &  90& elliptical  & cluster     &3    &   80 &   7200 &38.1&11.6&310.3\\
NGC 6010     &  S0-a&   0.4 &    12.95   &  90&             &             &1    &   15 &  13800 &4.2&0.9&148.8\\
NGC 6829     &  Sb  &   3.0 &    15.09   &  90&             &             &2    &  121 &  12800 &7.1&1.4&211.9\\
NGC 7183     &  S0 &  $-0.7$ &  13.14   &  90& close to box&             &5    &   77 &   7200 &2.5&0.5&209.9\\
NGC 7264     &  Sb  &   3.1 &    14.58   &  90& elliptical  &             &2    &  147 &  13000 &1.2&0.4&248.0\\
NGC 7332     &  S0  &  $-2.0$ &  12.56   &  90& close to box&             &1    &   65 &  15000 &7.5&1.1&121.3\\
NGC 7396     &  Sa  &   1.0 &    13.85   &  60& elliptical  &             &2    &   13 &  13000 &25.2&1.6&360.7\\
NGC 7703     &  S0  &  $-2.0$ &  14.41   &  90& elliptical  & cluster     &2    &   57 &   6400 &4.8&2.0&159.5\\
NGC 7814     &  SAab &   2.0 &    11.68   &  67& elliptical  &             &1    &   45 &  18400 &33.5&2.0&231.0\\
IC 1711      &  Sb  &   3.0 &    14.53   &  90& elliptical  &             &2    &  133 &   6200 &2.1&0.5&173.4\\
IC 1970      &  SAb  &   3.1 &    13.36   &  88& elliptical  &             &4    &   75 &   7200 &8.1&1.0&118.9\\
IC 2531      &  Sb  &   4.9 &    12.88   &  90& peanut      & cluster     &3    &   75 &   7200 &2.3&0.3&228.1\\
IC 5176      &  SABbc&   4.4 &    13.02   &  90& elliptical  & companions? &5    &  151 &   7200 &4.6&0.3&164.2\\
IC 5264      &  Sab &   2.3 &    13.58   &  90& elliptical  & group       &4,5  &   82 &  12600 &2.8&0.1&230.6\\
UGC 10043    &  Sbc &   4.1 &    15.17   &  90&             & cluster     &2    &   61 &  12200 &1.0&0.3&143.6\\
UGC 11552    &  Sab &   2.0 &    15.68   &  75&             &             &5    &   16 &   7200 &11.7&0.6&243.0\\
UGC 11587    &  S0  &  $-1.9$ &  14.52   &  90&             &             &1    &   76 &  15000 &5.7&0.7& \\
ESO 079-003  &  SBb &   3.1 &    13.91   &  90& close to box&             &4,5  &  134 &  10800 &5.9&2.0&192.2\\
ESO 234-053  &  S0  &  $-1.9$ &  14.32   &  90&             &             &5    &   81 &   5400 &1.7&0.1& \\
ESO 311-012  &  S0-a&   0.1 &    12.38   &  90& box         & isolated    &3    &   14 &   7200 &7.4&1.3&95.9\\
ESO 443-042  &  Sb  &   3.1 &    13.85   &  90& peanut      & companions? &3    &  130 &  10800 &3.8&0.7&166.3\\
ESO 512-012  &  Sbc  &   3.2 &    14.42   &  90& peanut      &             &3    &  113 &   5400 &1.4&0.3&224.9\\
\hline
\end{tabular}
\begin{list}{}{}

\item[$^{\rm a}$] From L\"utticke et al.\ (2000).

\item[$^{\rm b}$] Run number in which each galaxy was observed.

\item[$^{\rm c}$] Position angle of the long slit.

\item[$^{\rm d}$] Total exposure time (in seconds).

\item[$^{\rm e}$] Effective radius (in arcseconds) of the bulge,
computed as explained in Section~\ref{s:reduction}.

\item[$^{\rm f}$] Maximum rotational velocity (in km s$^{-1}$)
along the disk major axis, corrected from inclination when
necessary.

\item[Hubble type $T$, total apparent magnitude $B_{\rm T}$,
galaxy inclination $i$ and $V_{\rm max}$ have been taken from
HyperLeda (http://leda.univ-lyon1.fr/).]

\end{list}

\end{table*}

\section{Spectroscopic observations}  \label{s:obs}

Spectroscopic observations of the galaxy sample were obtained during five
observing runs at three different telescopes:

The northern galaxies were observed at the 2.5-m Isaac Newton Telescope
(INT) of the Isaac Newton Group of telescopes on the island of La Palma
(Spain). For the southern galaxies we used the 3.6-m ESO telescope and 3.6-m
New Technology Telescope (NTT), both at ESO La Silla observatory (Chile).

The main parameters of the instrument configurations and other details of the
observing runs are listed in Table~\ref{truns}.

Using  slit  widths  from  1.0   to   2.0 arcseconds, the    different
spectrographs  allowed us   to  obtain spectra   at a   resolution  of
typically 4~\AA\  (FWHM, as measured   from the widths of  arc lines),
except for  run~3 (the ESO 3.6-m run),  where a resolution of 8.3~\AA\
was attained.

The spectral range, from around $\lambda3900$\AA\ to $\lambda5500$\AA\
(with the exception of the ESO NTT runs, in which the spectra extended
redward to $\lambda6650$\AA) was chosen  to include the main prominent
spectral  features of the  blue-yellow   spectra of the galaxies,   in
particular most of the Lick/IDS line-strength indices.

Table~\ref{tgalaxies} includes some relevant observational parameters for the
galaxy sample. The spectrograph slit was oriented along the minor axis of the
bulges.

Thus, effects of contamination by disk light are only expected for the
very central regions of the sample galaxies.

Exposure times ranged from 1.5 to 5 hours per galaxy, providing high
signal-to-noise ($S/N$) spectra out to the outer regions of most bulges.

We obtained reliable  spectra   out to the effective  radius  ($r_{\rm
  eff}$) for  every galaxy  in our  sample.  Furthermore, our  spectra
reached $S/N>10$ (per  \AA\ and per arcsec  in the  spatial direction)
beyond   $2 r_{\rm  eff}$    for 80\%  of  the   sample galaxies. Some
observations of nearby  bulges were especially deep, providing  useful
spectra out to several kpc from the centres of the bulges (e.g., $\sim
7$ kpc for NGC~5084 and $\sim 4$ kpc for NGC~7814).

In the central regions we obtained spectra with S/N ranging from 20 to
240, depending, among other factors,  on the obscuration by dust lanes
within the galaxy disk.

In each observing run we also acquired spectra of a number of template
stars from the Lick/IDS library (Worthey et al.\ 1994) for calibration
purposes (see below)  as well as several  spectrophotometric standards
to calibrate the spectra in flux.

\begin{table*}[t]
\centering
\caption{Instrumental details of the observing runs.}
\label{truns}
\begin{tabular}{cccccccc} \hline \hline
 Run  &  Dates & Telescope & Instrument &  Dispersion & Spectral range &
Slit width & FWHM$^{\rm a}$ \\
  & & & &  \AA/pixel &  \AA &  arcsec &  \AA \\ \hline
1 & 13/08/96--16/08/96 & INT 2.5~m & IDS   & 1.6 & 3908--5535 & 2.0 & 3.9 \\
2 & 31/08/97--03/09/97 & INT 2.5~m & IDS   & 1.6 & 3908--5535 & 2.0 & 4.3 \\
3 & 08/03/97--11/03/97 & ESO 3.6~m & EFOSC & 3.4 & 3776--5504 & 1.5 & 8.3 \\
4 & 23/09/97--25/09/97 & NTT 3.6~m & EMMI  & 1.3 & 3971--6649 & 1.0 & 3.7 \\
5 & 23/08/98--26/08/98 & NTT 3.6~m & EMMI  & 1.3 & 3985--6649 & 1.5 & 4.9 \\
\hline
\end{tabular}

\begin{list}{}{}
\item[$^{\rm a}$] Spectral resolution as measured from arc lines.
\end{list}

\end{table*}

\section{Data reduction}
\label{s:reduction}

The     reduction    of   the   data     was     performed with    the
\reduceme\footnote{http://www.ucm.es/info/Astrof/software/reduceme/reduceme.html}\
package (Cardiel 1999). We carried  out a standard reduction procedure
for  spectroscopic data:   bias   and  dark subtraction,  cosmic   ray
cleaning, flat-fielding (using   observations of   tungsten lamps  and
twilight skies   to  correct from high  and   low frequency variations
respectively),     C-distortion    correction, wavelength calibration,
S-distortion   correction and    recentering  of    the spectra,   sky
subtraction,   atmospheric  and interstellar  extinction   correction,
relative flux  calibration, and spectrum  extraction.  \reduceme\ is a
package
%PG%
specifically    written    to     reduce   long-slit     spectroscopic
observation. Its main advantage is  that, for each observed data frame
(including calibration ones), it creates  an associated error frame at
the beginning of the reduction  procedure.  From this point, error and
data   frames   are processed    in  parallel,  translating    all the
uncertainties in  the manipulation of the data  into the error frames,
and

hence providing very accurate estimates of the random errors

associated with the final spectra (see Cardiel et al.\ 1998).

Although most of the reduction steps were performed using standard
procedures for long-slit reduction, we give below some comments on steps of
particular importance:

{\it (i) Wavelength calibration}.

Spectra were converted to a linear wavelength scale using
about 90 arc lines 
fitted by 5th to 7th-order polynomials. The rms dispersion of the fitted
position of arc lines was typically 0.25--0.30~\AA.

{\it (ii) Spectral rectification and centering}.

As  usual with  CCD  spectrographs,    the  galaxy  spectra were   not
completely perfectly aligned with  the detector rows.  To correct for
this effect (which is crucial when measuring line-strength gradients),
we fitted a  Cauchy  function to the number   of counts in  a  spatial
interval  around the location   of  maximum counts  for each  spectral
resolution element,  hence determining  the   position of the   galaxy
centre   as function of  wavelength.    In galaxies  with dust   lanes
obscuring the light of the central region of  the bulge, these central
parts were  masked  when fitting the luminosity  profiles. Experiments
with galaxies without dust lanes showed  that we could still trace the
position of the galaxy centre. The resulting  map was then fitted by a
low-order polynomial, which   was   in turn used   to  straighten  the
spectra. As a result   of this process,  the  central spectrum of  the
galaxy is moved to the centre of a detector row
so that symmetric spectra on each side of that row correspond to
the same galactocentric radius.

{\it (iii) Sky subtraction}.

Accurate sky subtraction is critical for studies like this

since we aim to analyze spectra  at light levels corresponding to only
a few  per cent of  the sky signal. We refer  the reader to Cardiel et
al.\ (1995) for  a description of  the sky subtraction effects  on the
measurements  of  line-strength   indices.   For each  of   our galaxy
observations we  generated a sky image  by fitting, for  each pixel in
the  wavelength direction,  a low-order  polynomial  using the regions
close to the slit ends. In most cases, the  galaxies fill only a small
region  of the   slit,  so  this synthetic   sky  image is   free from
contamination from the galaxy light. However, this is not the case for
some  of   the larger  galaxies  of  the  sample   (like  NGC~7814 and
NGC~5084).   In  these cases,   we  fitted  de  Vaucouleurs' $r^{1/4}$
profiles to the surface brightness distribution of the galaxies
along the slit,

estimating the relative contribution of the galaxy to the regions from
where the  sky was extracted. We then  `decontaminated' the sky frames
by subtracting scaled and averaged  galaxy spectra. In order to assess
the quality of the sky subtraction we looked  for residuals due to sky
emission lines, making sure that they were completely removed, even in
the outer parts of the more extended galaxies.

{\it (iv) Measurement of effective radii}.

Effective  radii  for all the bulges  of  the sample  were measured by
fitting de Vaucouleurs'   profiles to  the radial  surface  brightness
profiles  determined by  collapsing    the spectra in   the wavelength
direction. The central regions, affected in  many cases by dust lanes,
were removed from the fit. In any case, we never used spectra at radii
below typically 2 arcsec. On the other hand, the fits were quite
insensitive to the outer radius cutoff, although deviations
of the  luminosity  profiles from a  straight  line in the logarithmic
plots were sometimes observed. The derived  effective radii are listed
in Table~\ref{tgalaxies}. The  associated errors, in the  last column,
reflect the sensitivity of the derived radii  to the particular choice
of the  region to be  fitted.  Note that we  are not assuming that the
bulges do follow a de Vaucouleurs luminosity profile
\citep*[a Sersic profile  with an exponent $n$ closer  to 1
provides a better  fit in many     cases, see][]{andr+95}.
However, the derived effective radii provide us with a well
defined parameter to express the line-strength gradients
(Paper~II) on an homogeneous spatial scale.

{\it (v) Flux calibration}.

Relative flux calibration of the spectra was achieved using
exposures of Massey et al.\ (1988), Oke (1990) and Hamuy et al.\
(1994) spectrophotometric standards. In each run we took several
repeated observations of 4, 5, 4, 3 and 3 stars (for runs 1 to 5
respectively). In some cases we obtained exposures at different
zenithal distances to check the atmospheric extinction correction.
All the calibration curves of each run were averaged and the flux
calibration errors were estimated by the differences among the
indices measured with different curves.

{\it (vi) Extraction of spectra}.

Once the galaxy frames were fully reduced, we  added all the frames of
the same galaxies and extracted individual spectra along the slit. The
availability of reliable error frames allowed us to compute the number
of  spatial bins to  be coadded  to  guarantee a minimum  $S/N$ in the
final spectra. Using the    prescriptions given  in Cardiel  et   al.\
(1998), we chose a lower $S/N$ cutoff of 15 (per \AA; mean value along
the full spectral range), which ensures a typical error in most of the
atomic   Lick indices below 20\%.  For  the  outer spatial regions, we
coadded some further spectra to get a minimum $S/N$ of 10, which
is enough to measure molecular indices (like Mg$_2$) with a relative
uncertainty below 10\%, even though it does lead to significant errors
in the atomic indices.

\section{Dynamical parameters}
\label{s:dyn}

Radial velocities and velocity dispersions for each spectrum along
the bulge radii were measured using a  dedicated program within
\reduceme\ which incorporates the {\sc MOVEL} and {\sc OPTEMA}
algorithms devised by Gonz\'{a}lez (1993). The {\sc MOVEL}
procedure is a refined version of the classic  Fourier quotient
method (Sargent  et  al.\ 1977). The basic improvement over the
standard procedure is that, starting from a first    guess of
$\gamma$  (mean    line-strength), $V$  (radial velocity)  and
$\sigma$ (velocity    dispersion), it creates a   model galaxy
which is  processed in  parallel to  the  galaxy spectrum. The
differences between   the input and   output  parameters of  the
model spectrum are  then  used to correct   the  galaxy power
spectrum from systematic effects  in the handling of  data in the
Fourier space. The {\sc OPTEMA}  algorithm allowed  us  to
overcome  the typical template mismatch problem.  In order to do
this, a number of stellar spectra of different spectral types and
luminosity classes is fed into the program. In our case, we
combined our template stars to   construct  representative input
spectra  of  the following types: A3-5V, G0-2V, G5V, K0-1V,
G8-K0III, K3III, K5-7III, and M3III. The  algorithm   then
constructs an   optimal   template as  a linear combination  of
the  above  template   spectra which  minimizes   the residuals
between the   galaxy  spectrum and the   broadened template.
Meanwhile,  the dynamical parameters  are derived as  part of the
same minimization process. Our tests showed that  if a single
template star is used  to measure  the  velocity dispersion
(instead of  the above combination), systematic errors as large as
10\% can be introduced.

\begin{figure}
\resizebox{1.0\hsize}{!}{\includegraphics[angle=-90]{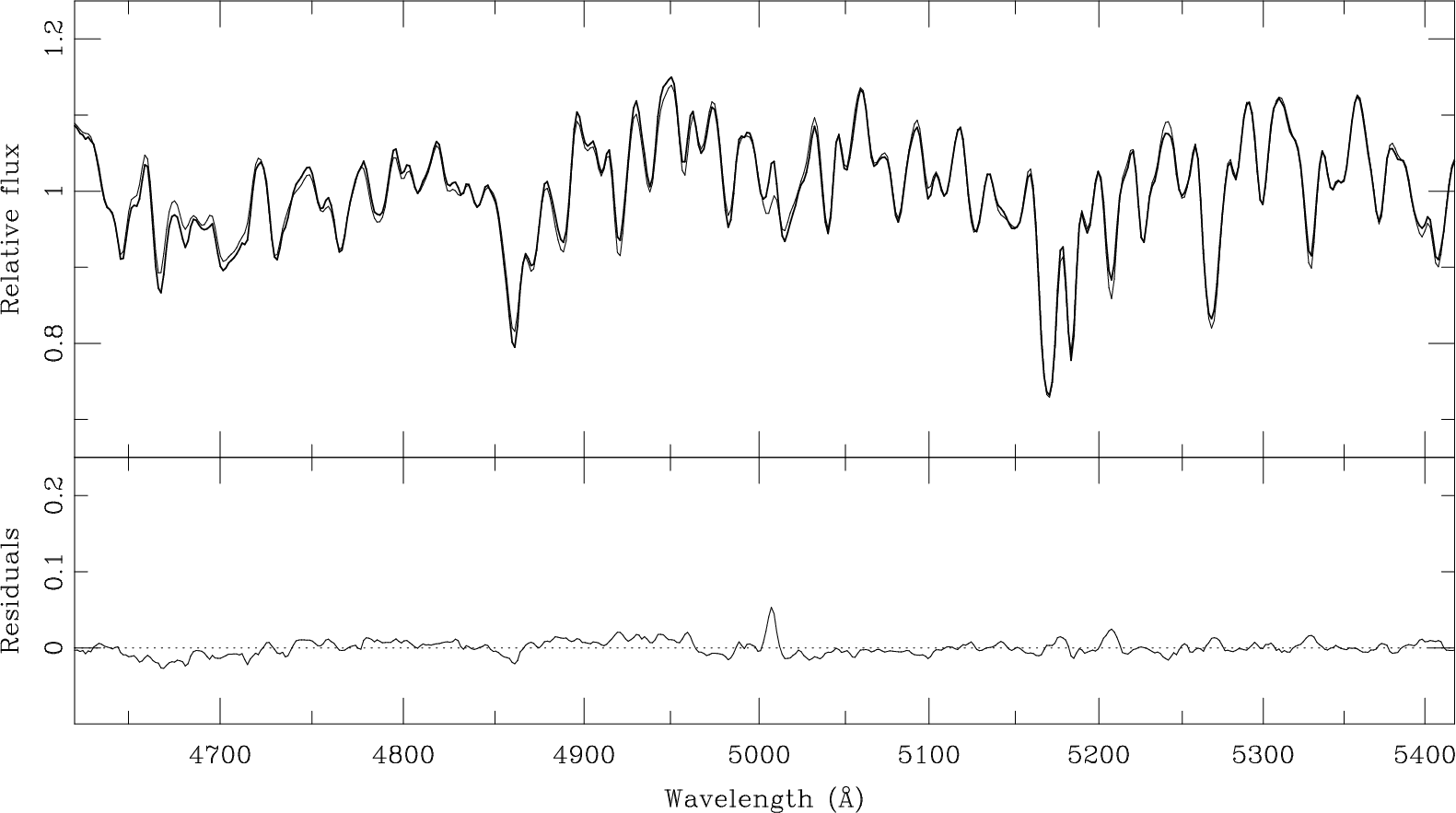}}
\resizebox{1.0\hsize}{!}{\includegraphics[angle=-90]{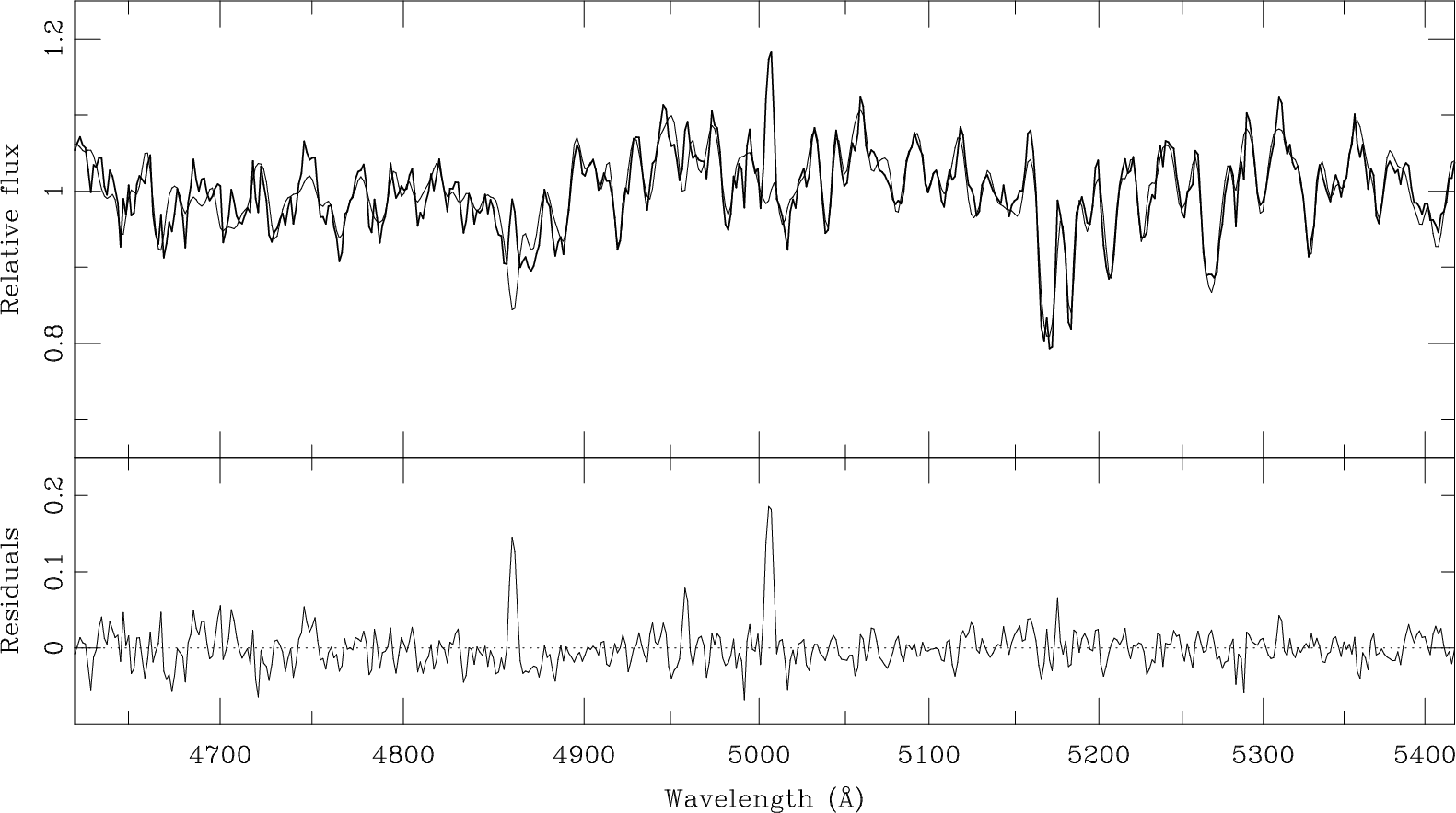}}
 \caption{Illustration of the velocity dispersion determinations using optimal templates for
 NGC~7332 (upper plot) and UGC~10043 (lower plot). For each plot, the upper
 panels show the observed galaxy spectrum (thick line) and the corrected
 optimal template spectrum (thin line). The residuals of the fits are shown in
 the lower panels.}
 \label{fitemplate}
\end{figure}

Figure~\ref{fitemplate} illustrates  the result of this  procedure for
two  galaxies (NGC~7332 and UGC~10043), showing,  in the upper panels,
the fits between the observed central spectra of  the galaxies and the
corresponding optimal templates corrected  with the  derived dynamical
parameters. The lower panels show the residual of  the fits. Note that
emission lines, although almost hidden in  the original spectrum (like
the \OIII$\lambda$5007 line  for NGC~7332) are clearly detected  after
subtracting   the template (once an    emission line is detected,  the
corresponding wavelength  region is not used  for  the minimization of
the residuals). In fact, the optimal templates derived for each galaxy
were later  used to  mask  the  regions possibly  containing  emission
lines.

Realistic  errors  in the   derived parameters  (radial  velocity
and velocity    dispersion)   were  computed    by performing
Monte-Carlo simulations, repeating the whole process  (including
the derivation of the    optimal template) for  a large   number
of simulated galaxy spectra created using the error spectrum
obtained during the reduction process. We checked that,  in order
to measure velocity dispersion of the order of the spectral
resolution  ($\sim 100$~km~s$^{-1}$ for runs 1, 2, 4 and 5) with a
relative error below 20\% we needed spectra with $S/N$ (per \AA)
larger than 15.
This coincides with the cutoff value for the spectral extraction
extraction to measure the atomic indices.

Once an  optimal template is derived  for the central or maximum
$S/N$ (if the centre is  affected by dust  lanes) spectrum, the
procedure is repeated for  all the spectra along the  bulge radii,
deriving radial profiles  for  the   radial velocity and
velocity  dispersion. Note that the template derived for the
central spectrum is used for the analysis of the rest of the
spectra and that, therefore, that we are not taking into account
the possible effect of population gradients in the derived sigma
profiles. However, we preferred to introduce this small inaccuracy
than to artificially increase the scatter of the parameters
profiles due to unreal changes in the derived linear combinations
of templates at each radius. Note also that any radial change in
mean line-strength (i.e. $\gamma$) or continuum shape is properly
corrected from in the MOVEL algorithm.

The dynamical parameters, and  their  corresponding errors,  at
each galactocentric radius are listed in Table~\ref{tsigmas}.
Since the slit was oriented along  the minor axes  of the bulges,
no rotation  along  the slit was expected. To check this, in
Table~\ref{tsigmas} we include the maximum rotational   velocity
(computed   as  an error-weighted   mean of the, generally two,
spectra at   each side of  the  galaxy centre with  the highest
rotational velocity) and its error.  It is  apparent from the
Table that most bulges do not show any  significant rotation along
the minor  axes.  For some galaxies,  however,  we  observed some
hints of rotation  or peculiar $V_r$ profiles  in the central
regions, probably due to slight misalignments of the slit with
respect to the minor axis direction or the bulge centre.

For galaxies with  no obvious dust  lanes in the central  regions, the
table   also includes  the   central  velocity  dispersion $\sigma_0$,
measured  in an aperture  of 4 arcsec times  the slit width centred on
the  galaxy     centre.   We  also list     mean  velocity dispersions
($\left<\sigma\right>$)  defined  as the error-weighted  mean velocity
dispersion  of all  the extracted spectra   along the radius (which is
very close to the velocity dispersion that  is derived when adding all
the spectra along the slit).

\begin{figure*}
\centerline{
\resizebox{0.9\hsize}{!}{\includegraphics{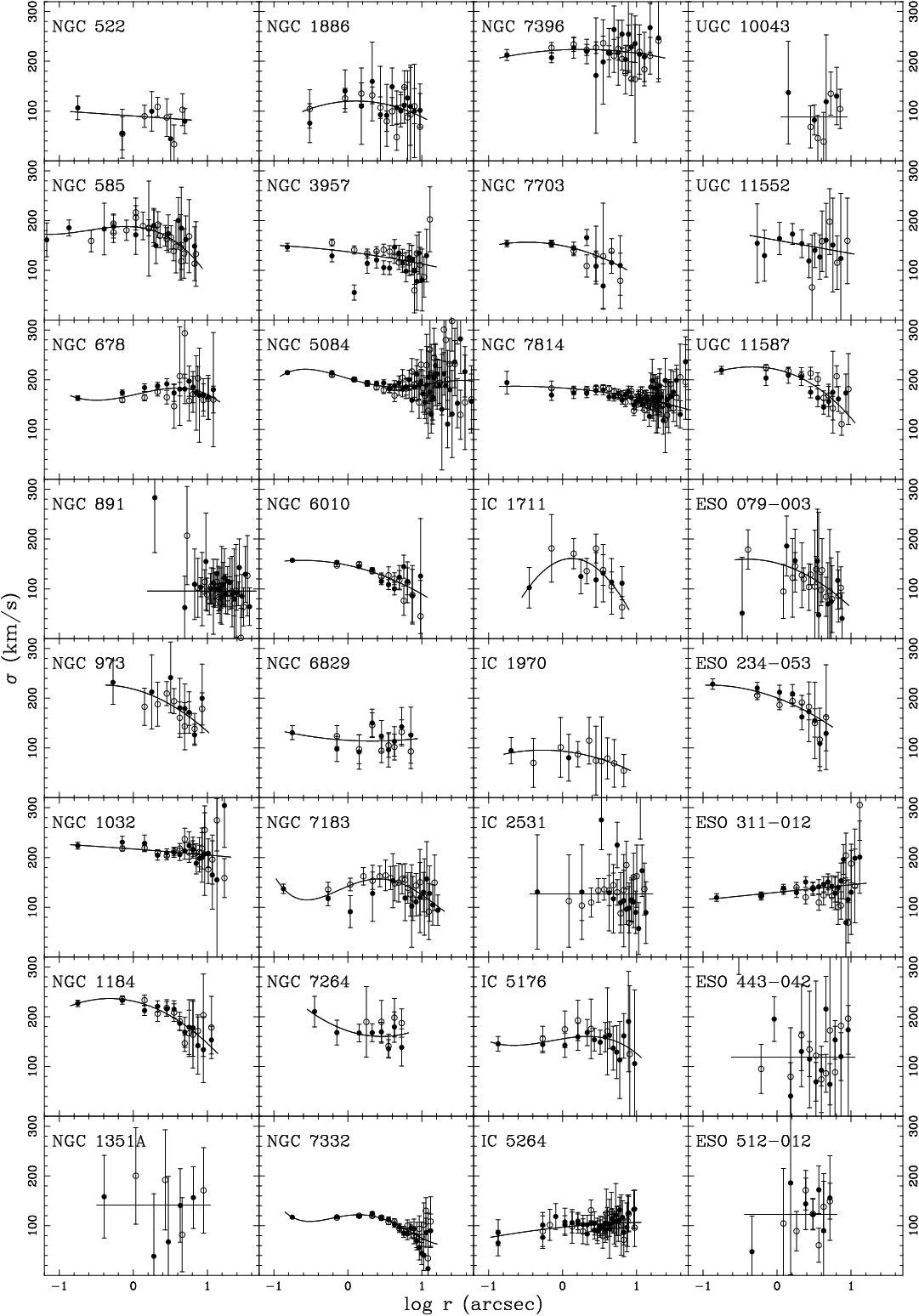}}
} \caption{Velocity dispersion profiles for the galaxy sample. Filled
and open symbols refer to spectra of the bulges at either side of the
galaxy centres. The plotted lines represent polynomial fits to the
profiles. See text for details.}  \label{panelsigma}
\end{figure*}

The  derived velocity dispersion  profiles  for the galaxy sample  are
plotted in  Figure~\ref{panelsigma}. It can  be  seen that, typically,
the velocity dispersion decreases with galactocentric radius, although
there   are different behaviors and  some   galaxies exhibit a  large
scatter.  It should  be noted  that the main  motivation for measuring
velocity dispersion profiles in this   paper is to have the  necessary
information to carry out accurate  corrections to the Lick/IDS indices
(detailed further below).

Since line-strength indices depend on the spectral resolution, all
spectra along the slit must be corrected to the same velocity
dispersion before measuring the indices so as to avoid the
introduction of spurious line-strength gradients. However, if we
used the nominal velocity dispersion values obtained for each
individual spectrum to perform this correction, the errors
associated with the $\sigma$ measurement would be introduced in
the measured indices.

To avoid this, we smoothed the velocity dispersion profiles by fitting
low-order polynomials  to the  data,  and we used  the  predictions of
these polynomials  for correcting the indices  to the desired velocity
dispersions. Note that we are  not  pretending to assign any  physical
meaning to these  polynomial     radial profiles; they only  have    a
calibration     purpose.     The  column    $\sigma_{\rm    max}$   in
Table~\ref{tsigmas} gives the maximum velocity dispersion indicated by
the polynomials in the observed radial range.

Finally, we compared  the derived values  with the available data from
the literature (last columns  of Table~\ref{tsigmas}) to check for any
systematic offset in  the derived   velocity dispersion. To   minimize
aperture differences, these reference data should be compared with our
central    ($\sigma_0$)   values     when    available,     or    with
$\left<\sigma\right>$  otherwise. It can  be seen that  we do not have
any systematic offset relative to the scarce previous data.

\begin{table*}[t]
\centering \caption{Dynamical parameters of the bulge sample.}
\label{tsigmas}
\begin{tabular}{lrrrrrrrrrrc} \hline \hline
Galaxy & \multicolumn{1}{c}{$V_{\rm r}^{\rm a}$} & \multicolumn{1}{c}{$\Delta
V_{\rm r}$} & \multicolumn{1}{c}{$V_{\rm rot}^{\rm b}$} &
\multicolumn{1}{c}{$\Delta V_{\rm rot}$} & \multicolumn{1}{c}{$\sigma_0^{\rm
c}$} & \multicolumn{1}{c}{$\Delta\sigma_0$} & \multicolumn{1}{c}{$<\sigma>^{\rm
d}$} & \multicolumn{1}{c}{$\Delta <\sigma>$} & \multicolumn{1}{c}{$\sigma_{\rm
max}^{\rm e}$} & \multicolumn{1}{c}{$\sigma_{\rm ref}$} & Ref$^{\rm f}$\\
\hline
NGC 522    & 2719.0& 12.5& 12.1&  4.9&  82.1& 9.3&  88.2&  8.8&  99&    &   \\
NGC 585    & 5358.4&  6.8&  5.5&  4.5&      &    & 174.3&  4.7& 188&    &   \\
NGC 678    & 2814.9& 15.0& 13.3&  6.3& 163.4& 2.1& 170.7&  4.1& 183&    &   \\
NGC 891    &  548.2& 12.3&  1.9&  2.2&      &    &  95.4&  5.1&  95&  72&(2)\\
NGC 973    & 4081.0&  9.6&  1.5&  6.0&      &    & 173.3&  9.3& 226&    &   \\
NGC 1032   & 2702.9&  7.7& 16.3&  9.4&      &    & 214.2&  2.8& 225&    &   \\
NGC 1184   & 2278.2&  8.0&  5.1&  3.1& 230.3& 3.0& 219.6&  7.6& 234& 229&(3)\\
NGC 1351A  & 1329.3& 17.6&  6.1& 15.4&      &    & 141.4& 28.5& 141&    &   \\
NGC 1886   & 1689.8& 15.1&  8.0&  3.3&      &    & 111.0&  6.5& 120&    &   \\
NGC 3957   & 1589.7& 10.2&  1.6&  2.2&      &    & 131.4&  5.3& 149&    &   \\
NGC 5084   & 1671.6& 16.8& 10.5&  9.5& 207.6& 1.7& 200.0&  4.0& 214& 215&(1)\\
NGC 6010   & 2058.5&  3.3&  2.9&  3.3& 148.4& 1.2& 146.7&  4.9& 157& 144&(4)\\
NGC 6829   & 3303.3& 11.0&  1.2&  5.6&      &    & 118.1&  6.3& 130&    &   \\
NGC 7183   & 2594.9& 12.8&  2.2&  5.0&      &    & 138.8&  5.3& 157&    &   \\
NGC 7264   & 4210.1& 14.4&  6.9&  4.6&      &    & 165.6&  6.9& 205&    &   \\
NGC 7332   & 1206.3&  2.2&  2.6&  2.8& 117.2& 0.6& 116.1&  2.8& 122& 134&(1)\\
NGC 7396   & 4882.5& 12.2&  8.9&  6.6&      &    & 220.6&  4.1& 224&    &   \\
NGC 7703   & 3970.5&  7.1&  7.1&  4.8& 149.9& 3.2& 148.5&  5.9& 154& 166&(3)\\
NGC 7814   & 1051.4&  6.9&  4.8&  4.5&      &    & 170.3&  2.4& 187& 172&(1)\\
IC 1711    & 2785.2& 11.8&  9.0&  5.2&      &    & 122.3& 12.2& 161&    &   \\
IC 1970    & 1300.8&  7.2& 11.6& 10.6&      &    &  82.8& 12.1&  95&    &   \\
IC 2531    & 2386.8& 11.7&  3.8&  4.1&      &    & 126.6&  8.7& 127&    &   \\
IC 5176    & 1739.3& 10.5&  4.1&  4.3&      &    & 152.7&  6.4& 160&    &   \\
IC 5264    & 2025.1&  6.3&  2.6&  3.6&      &    & 101.1&  3.4& 106&    &   \\
UGC 10043  & 2180.3& 16.0&  8.2&  6.4&      &    &  88.5& 15.7&  89&    &   \\
UGC 11552  & 4479.5& 11.1&  2.7& 11.3&      &    & 151.3& 10.7& 167&    &   \\
UGC 11587  & 4481.5&  7.8&  2.3&  4.9& 213.0& 2.9& 203.4&  8.5& 224&    &   \\
ESO 079-003& 2613.5&  8.1&  1.5&  6.8&      &    & 120.7&  8.5& 160&    &   \\
ESO 234-053& 6232.9& 16.0&  0.3&  4.9& 204.9& 3.1& 200.4&  7.6& 226&    &   \\
ESO 311-012& 1106.8& 10.7&  3.2&  5.3& 124.2& 3.5& 132.1&  4.0& 147&    &   \\
ESO 443-042& 2815.8& 24.8&  8.7&  9.6&      &    & 118.6& 14.9& 119&    &   \\
ESO 512-012& 3313.4& 19.0& 13.0&  6.5&      &    & 122.8& 12.4& 123&    &   \\
\hline
\end{tabular}

\begin{list}{}{}
\item[$^{\rm a}$] Mean radial velocity (in km s$^{-1}$).

\item[$^{\rm b}$] Maximum rotational velocity (see the text).

\item[$^{\rm c}$] Central velocity dispersion.

\item[$^{\rm d}$] Mean velocity dispersion.

\item[$^{\rm e}$] Maximum velocity dispersion (see the text).

\item[$^{\rm f}$] References are: (1) Average value quoted in {\sc HYPERLEDA};
(2) Bottema et al.\ (1991); (3) Di Nella et al.\ (1995), and (4)
Falc\'{o}n-Barroso et al.\ (2002).

\end{list}

\end{table*}

\section{Line-strength indices}   \label{s:indices}

Line-strength indices in the Lick/IDS system (e.g. Worthey et al.\ 1994; Worthey
\& Ottaviani 1997) were measured for all the extracted spectra along the
measured extent of the bulge. We measured all the indices from H$\delta$ to
Fe5406. In the case of
runs 4 and 5 (cf.\ Table~\ref{truns}), the extended spectral range allowed us
to include also the redder
Lick indices (up to the TiO indices). The next subsections explain all the
procedures to derive reliable errors in the line-strengths and to transform
them to the spectrophotometric Lick/IDS system. All these steps were possible
thanks to the inclusion of a number of stars from the Lick/IDS library in the
different observing runs. In particular, we observed 40, 40, 19, 4 and 6 Lick
standard stars (covering a range of spectral types and luminosity classes) in
runs
from 1 to 5 respectively.

\subsection{Error analysis}

A first estimation of errors in the line strengths was accomplished by
taking into account three different error  sources: {\it (i)\/} photon
noise (the errors were calculated  using the variance spectra computed
as part of the reduction process),  {\it (ii)\/} flux calibration (the
estimation was  based on the  comparison of the indices measured using
the different flux calibration curves obtained in each observing run),
and {\it (iii)\/}uncertainties in  the wavelength scale (due to errors
in the  wavelength calibration and  in  the measurement of  the radial
velocity).

Except for the spectra with the highest $S/N$, the dominant error source was
the first one.

In order  to check and  refine  the error  analysis,  we analyzed  the
line-strengths obtained for:  {\it (i)\/} repeated observations of the
same stars within   each observing run,  {\it (ii)\/}  stars in common
among different observing runs (e.g. all the 40 Lick stars observed in
run 1 were repeated in run 2), and {\it (iii)\/} separate observations
of the same  galaxies within each  run (typically the observation of a
galaxy was splitted into 4-12  different exposures). Using this  data,
we computed  residual errors as  the quadratic differences between the
measured r.m.s.\  dispersions  and the expected  total  error from the
three sources of  uncertainty  quoted in the previous   paragraph. The
resulting residual  errors are listed  in  Table~\ref{indres} for each
index and observing run. They are relatively  modest but were included
anyway in the final uncertainties.

\subsection{Emission-line corrections}

Some line-strength indices, in particular H$\beta$, Fe5015 and Mgb, are
potentially affected by emission lines (e.g., Goudfrooij \& Emsellem
1996). Most of our galaxies exhibit non-negligible nebular emission, as
judged from their [O\,{\sc iii}]$\lambda5007$ line.

Usually, the emission is concentrated towards the bulge centres, where
the spectra  are heavily contaminated by  the  disk light (H\,{\sc ii}
regions can  project onto the   slit). However, in some special  cases
(e.g., NGC~1886 and  NGC~7332), line emission can  be detected even in
the outer part of the bulge, where any  contamination by disk light is
negligible. For instance, in NGC~7332,  the line emission increases to
$EW($[O\,{\sc   iii}]$\lambda5007)\simeq4$    \AA\  at  a  distance of
$\sim$\,16 arcsec from the centre.

The correction of H$\beta$ from the emission contribution is a
troublesome task. The usual procedure (see eg. Gonz\'alez 1993,
Trager et al.\ 2000) relies on using the equivalent width of
[O\,{\sc iii}] to apply an empirical correction. In this paper,
given the uncertainties of that correction and the fact that the
emission is relatively strong for many of the galaxies, we have
decided not to correct H$\beta$ and, therefore, not to use it in
our analysis in the cases where any [O\,{\sc iii}]$\lambda5007$
emission was detected (in particular when its measured equivalent
width was larger than its error).

We determined equivalent widths of [O\,{\sc iii}]$\lambda5007$ after
subtracting from each galaxy spectrum an emission-free template spectrum
constructed from the broadened optimal template derived for that galaxy
during the velocity dispersion measurement (see Figure~\ref{fitemplate}).

In order  to measure   the Fe5015  and Mgb   indices in  spectra  with
emission lines,  we masked    the  spectral regions affected   by  the
[O\,{\sc iii}]$\lambda4959$, [O\,{\sc iii}]$\lambda5007$, and [N\,{\sc
i}]$\lambda5199$ lines  and   replaced  them with   the  corresponding
regions in the broadened optimal template spectrum.

\subsection{Broadening corrections}

As mentioned before in Section~\ref{s:dyn}, the Lick/IDS indices have to be
corrected for spectral broadening. In order to avoid

the introduction of artificial line-strength  gradients due to changes
in velocity dispersion along the slit, we broadened all the spectra of
a  given galaxy to  match the maximum velocity dispersion $\sigma_{\rm
max}$ derived in Section~\ref{s:dyn} by convolving  it with a Gaussian
of width  $(\sigma_{\rm max}^2-\sigma_r^2)^{1/2}$, where $\sigma_r$ is
the velocity dispersion at each  radius predicted by the corresponding
polynomial fit.

The conversion of the line indices to the resolution of the
Lick system (see Section~\ref{sec.lick}) was done as follows.
In the cases where the total resolution of the spectra (the
quadratic sum of instrumental resolution and the maximum velocity
dispersion for the galaxy) was below the Lick resolution, we
simply broadened the spectra by the quadratic difference. In other
case, we applied an empirical correction to the measured indices.
This correction was derived by broadening stellar
spectra by convolution with Gaussians having a range of $\sigma$ values.
Since this correction depends on the individual spectral type of
the star, we used instead the optimal templates obtained in the
velocity dispersion measurement procedure to derive the following
expression:

\begin{equation}\label{sigcorpol}
I(\sigma_0)=I(\sigma)\left(c_1+c_2\
\sigma/\sigma_0+c_3\left(\sigma/\sigma_0\right)^2\right) ,
\end{equation}

\noindent which can be used to convert an index $I$ measured at a
resolution $\sigma$ to any other resolution $\sigma_0$. The values
of the $c_i$ coefficients are listed in Table~\ref{sigcoefs} for
all the Lick indices. In principle, the exact values of the
coefficients depend on the particular optimal template used.
However, the spectral differences among them have a small effect
on the derived polynomials and the uncertainties introduced by
using the listed mean coefficients are well below the other
sources of errors.

\begin{table}[t]
\centering
\caption{Coefficients of the polynomials to correct Lick indices
for broadening effects.}
\label{sigcoefs}
\begin{tabular}{lrrr} \hline \hline
Index &  \multicolumn{1}{c}{$c_1$}  & \multicolumn{1}{c}{$c_2$} &
\multicolumn{1}{c}{$c_3$} \\ \hline
H$\delta_{\rm A}$ & 0.76129 & 0.10120 & 0.13751 \\
H$\delta_{\rm F}$ & 1.25993 &$-$0.51501 & 0.25508 \\
CN$_1$            & 0.70442 & 0.31941 &$-$0.02382 \\
CN$_2$            & 0.76492 &$-$0.00441 & 0.23949 \\
Ca4227            & 0.78022 &$-$0.28090 & 0.50067 \\
G4300             & 0.89152 & 0.10002 & 0.00847 \\
H$\gamma_{\rm A}$ & 1.11845 &$-$0.21248 & 0.09403 \\
H$\gamma_{\rm F}$ & 0.80429 & 0.25651 &$-$0.06080 \\
Fe4383            & 0.87985 &$-$0.00952 & 0.12967 \\
Ca4455            & 0.63849 & 0.09619 & 0.26532 \\
Fe4531            & 0.91332 &$-$0.01235 & 0.09903 \\
Fe4668            & 1.00545 &$-$0.12685 & 0.12140 \\
${\rm H}\beta$    & 0.99068 & 0.00563 & 0.00369 \\
Fe5015            & 0.78157 & 0.17915 & 0.03928 \\
Mg$_1$            & 0.97728 & 0.01066 & 0.01206 \\
Mg$_2$            & 0.98082 & 0.02199 &$-$0.00282 \\
Mgb               & 0.96450 &$-$0.07495 & 0.11045 \\
Fe5270            & 0.82533 & 0.12283 & 0.05184 \\
Fe5335            & 0.84324 &$-$0.08141 & 0.23817 \\
Fe5406            & 0.89017 &$-$0.14121 & 0.25104 \\
Fe5709            & 0.89437 & 0.01365 & 0.09198 \\
Fe5782            & 0.86475 &$-$0.14458 & 0.27983 \\
Na5895            & 0.94445 & 0.01617 & 0.03938 \\
TiO$_1$           & 1.00000 & 0.00000 & 0.00000 \\
TiO$_2$           & 1.00000 & 0.00000 & 0.00000 \\
\hline
\end{tabular}
\end{table}

\subsection{Conversion to the Lick/IDS system}
\label{sec.lick}

Transformation of  line-strengths    to the  Lick/IDS  system  is  not
straightforward.  Basically, there are two important effects that have
to  be   taken into account.   First,  the Lick/IDS  spectra   are not
flux-calibrated using a spectrophotometric standard star, but by using
a normalized  tungsten   lamp spectrum. Thus,   there  are significant
differences in the spectral  shapes when compared with flux-calibrated
spectra.  This effect causes
systematic offsets, especially for indices with broad bands like (e.g.) Mg$_2$.
Second, the spectral resolution of the Lick spectra is not constant with
wavelength, in particular it degrades considerably towards the blue end. This
has an important effect on narrow line indices, like many of the atomic
ones. For a comprehensive discussion on these effects, we refer the reader to
Worthey \& Ottaviani (1997).

To  transform our   data  to the   Lick/IDS  system  we performed  the
following steps. First,   we estimated the  resolution  at which  each
particular index  should be measured   by broadening our  large set of
stars in common with the Lick library to several line widths, in steps
of 25 km~s$^{-1}$.  We did not attempt to find precise resolutions for
each  index since these  depend on the  particular  set of stars being
used. Instead,  we estimated the  approximate line widths $\sigma_{\rm
L}$  which,  changing    softly  with    wavelength,  minimized    the
residuals. These  are given, for each index,  in the  second column of
Table~\ref{indres}. These  resolutions agree with the  rough estimates
given in Worthey \& Ottaviani (1997).

We then broadened all our stellar spectra to the above
$\sigma_{\rm L}$ resolutions and measured the line-strength
indices. The comparison of these measurements with those in the
original Lick/IDS spectra allowed us to derive mean offsets for
all the indices in each observing run. These additive offsets,
listed in Table~\ref{indres}, are mainly due to differences in
flux calibration between both systems, but also include a fine
tuning of the broadening corrections and systematic effects in
both our data (like uncertainties in the spectral resolution or
flux calibration) and the Lick/IDS data (like spectral
peculiarities which depend on the particular Lick/IDS run in which
each star was observed). Note that we obtain different offsets for
each observing run (although in most cases they are similar).
In fact,  the  comparison with Lick  data provides  us with  an anchor
system to correct for small offsets  among the different runs. It must
be noted  that  the offsets  for  run 3,   with a significantly  lower
spectral resolution, are  of the same order  than the  offsets for the
other  runs, which confirms that the  broadening corrections have been
properly done.

Figure~\ref{panelindices} shows a comparison of the line-strength
indices in the Lick/IDS spectra and the measurements in our sample
of stars after correcting to the Lick/IDS system. Error bars where
derived by combining the computed errors for our measurements and
the errors of the original Lick/IDS indices as taken from Worthey
et al.\ (1994). The last two columns of Table~\ref{indres} give
the r.m.s. standard deviation with respect to the zero offset, and
the standard deviation that should be expected from the computed
errors in the offsets. As it can be seen, the former are usually
larger than the latter, indicating that the errors in the $I({\rm
Lick})-I({\rm here})$ differences are somewhat underestimated.
Given all the tests that we have carried out with our data, we
think that this disagreement is most likely due to an
underestimate of the errors in the original Lick measurements.

\begin{table*}
\caption{Spectral resolutions, residual errors and offsets needed to transform
line-strength indices to the Lick/IDS system.} \label{indres} \centering
\begin{tabular}{l@{\hspace{7mm}}cp{0.3cm}llllp{0.3cm}
r@{.}lr@{.}lr@{.}lr@{.}l@{\hspace{1.2cm}}ll} \hline\hline
%\begin{tabular}{l@{\hspace{1cm}}cp{0.5cm}llllp{0.5cm}
%r@{.}lr@{.}lr@{.}lr@{.}l@{\hspace{1.5cm}}ll} \hline\hline
%
Index & $\sigma_{\rm L}^{\rm a}$ & & \multicolumn{4}{c}{Residual errors} & &
\multicolumn{8}{c}{Offsets} & \multicolumn{2}{c}{$\sigma^{\rm b}$} \\
& km~s$^{-1}$ & & run 1 & run 2 & run 3 & runs 4,5 & & \multicolumn{2}{c}{run
1} & \multicolumn{2}{c}{run 2} & \multicolumn{2}{c}{run 3} &
\multicolumn{2}{l}{runs 4,5} & \multicolumn{1}{c}{rms} &
\multicolumn{1}{c}{exp} \\ \hline

H$\delta_{\rm A}$ & 325  &  & 0.00  & 0.00  & 0.00  & 0.20     & & $-$0&34  &
$-$0&34  &    0&00  & $-$1&09  & 0.66 & 0.37\\
H$\delta_{\rm F}$ & 325  &  & 0.00  & 0.00  & 0.00  & 0.00     & & $-$0&12  &
$-$0&09  &    0&00  & $-$0&81  & 0.35 & 0.24\\
CN$_1$            & 325  &  & 0.005 & 0.005 & 0.005 & 0.008    & & $+$0&003 &
$+$0&003 & $-$0&009 &    0&000 & 0.020 & 0.015\\
CN$_2$            & 325  &  & 0.004 & 0.004 & 0.004 & 0.011    & & $+$0&008 &
$+$0&008 & $-$0&015 & $-$0&011 & 0.021 & 0.016\\
Ca4227            & 300  &  & 0.04  & 0.04  & 0.04  & 0.00     & &    0&00  &
0&00  & $-$0&08  &    0&00 & 0.23 & 0.17\\
G4300             & 300  &  & 0.13  & 0.13  & 0.13  & 0.00     & & $-$0&08  &
$-$0&08  &    0&00  &    0&00 & 0.38 & 0.32\\
H$\gamma_{\rm A}$ & 275  &  & 0.19  & 0.19  & 0.21  & 0.00     & & $+$0&41  &
$+$0&41  & $+$0&41  & $+$0&34 & 0.48 & 0.40\\
H$\gamma_{\rm F}$ & 275  &  & 0.04  & 0.00  & 0.00  & 0.00     & & $+$0&17  &
$+$0&17  &    0&00  & $+$0&10 & 0.22 & 0.19\\
Fe4383            & 250  &  & 0.22  & 0.21  & 0.21  & 0.09     & & $+$0&08  &
$+$0&08  &    0&00  & $-$0&36 & 0.53 & 0.45\\
Ca4455            & 250  &  & 0.04  & 0.04  & 0.04  & 0.00     & & $+$0&32  &
$+$0&32  &    0&00  & $+$0&28 & 0.20 & 0.20\\
Fe4531            & 250  &  & 0.10  & 0.10  & 0.10  & 0.18     & & $+$0&15  &
$+$0&15  & $-$0&32  &    0&00 & 0.35 & 0.31\\
Fe4668            & 250  &  & 0.15  & 0.19  & 0.15  & 0.57     & &    0&00  &
0&00  & $-$1&03  &    0&00 & 0.57 & 0.59\\
${\rm H}\beta$    & 225  &  & 0.10  & 0.10  & 0.10  & 0.04     & & $-$0&07  &
$-$0&07  &    0&00  & $-$0&14 & 0.22 & 0.20\\
Fe5015            & 200  &  & 0.15  & 0.19  & 0.15  & 0.21     & & $+$0&26  &
$+$0&26  & $-$0&11  &    0&00 & 0.64 & 0.40\\
Mg$_1$            & 200  &  & 0.004 & 0.004 & 0.004 & 0.007    & & $+$0&011 &
$+$0&003 & $+$0&016 & $+$0&009 & 0.011 & 0.010\\
Mg$_2$            & 200  &  & 0.006 & 0.005 & 0.003 & 0.005    & & $+$0&025 &
$+$0&020 & $+$0&021 & $+$0&016 & 0.012 & 0.010\\
Mgb               & 200  &  & 0.04  & 0.03  & 0.02  & 0.07     & &    0&00  &
0&00  & $-$0&08  &    0&00 & 0.31 & 0.17\\
Fe5270            & 200  &  & 0.00  & 0.00  & 0.00  & 0.11     & &    0&00  &
$+$0&03  &    0&00  &    0&00 & 0.23 & 0.16\\
Fe5335            & 200  &  & 0.02  & 0.02  & 0.02  & 0.16     & & $-$0&06  &
$-$0&06  &    0&00  & $+$0&18 & 0.26 & 0.16\\
Fe5406            & 200  &  & 0.04  & 0.04  & 0.04  & 0.03     & & $-$0&04  &
$-$0&04  & $-$0&19  &    0&00 & 0.23 & 0.14\\
Fe5709            & 200  &  &       &       &       & 0.00     & &
\multicolumn{6}{c}{}     & $-$0&08 & 0.14 & 0.14\\
Fe5782            & 200  &  &       &       &       & 0.00     & &
\multicolumn{6}{c}{}     & $+$0&14 & 0.10 & 0.13\\
Na5895            & 200  &  &       &       &       & 0.24     & &
\multicolumn{6}{c}{}     &    0&00 & 0.26 & 0.40\\
TiO$_1$           & 200  &  &       &       &       & 0.003    & &
\multicolumn{6}{c}{}    & $+$0&013 & 0.008 & 0.007\\
TiO$_2$           & 200  &  &       &       &       & 0.000    & &
\multicolumn{6}{c}{}    & $-$0&015 & 0.005 & 0.004\\
\hline
\end{tabular}
\begin{list}{}{}
\item[$^{\rm a}$] Spectral resolutions at which line-strength indices should be
measured to roughly match the original Lick/IDS system.

\item[$^{\rm b}$] Measured (rms) and expected (exp) standard deviations for the
comparison of indices with the original values of the Lick/IDS system.
\end{list}
\end{table*}

\begin{figure*}
\resizebox{1.0\hsize}{!}{\includegraphics{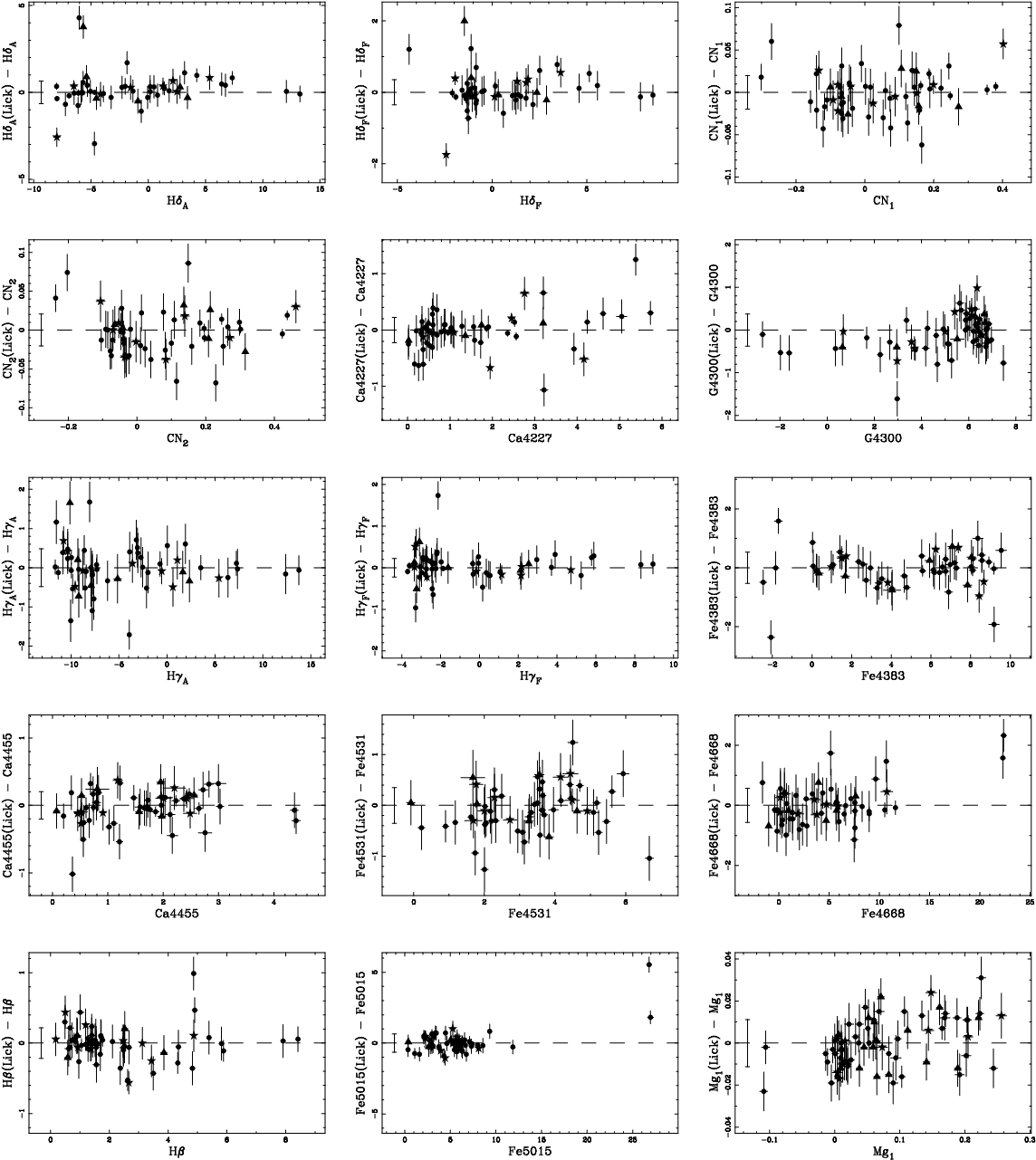}}
 \caption{Comparison of the line-strength indices in the Lick/IDS spectra and the
 measurements in our sample of stars after correcting to the Lick/IDS system.
 Different symbols refer to the different observing runs, using the following code:
 circles for runs 1 and 2; triangles for run 3, and stars for runs 4 and 5. In the
 case of the TiO indices, both TiO$_1$ are TiO$_2$ are plotted in the same panel, using
 closed and open symbols for both indices respectively. The length of the error bar near
 the left end of each panel is twice the r.m.s. standard deviation of each dataset with
 respect to the zero offset horizontal line.}
 \label{panelindices}
\end{figure*}

\begin{figure*}
\addtocounter{figure}{-1}
\resizebox{1.0\hsize}{!}{\includegraphics{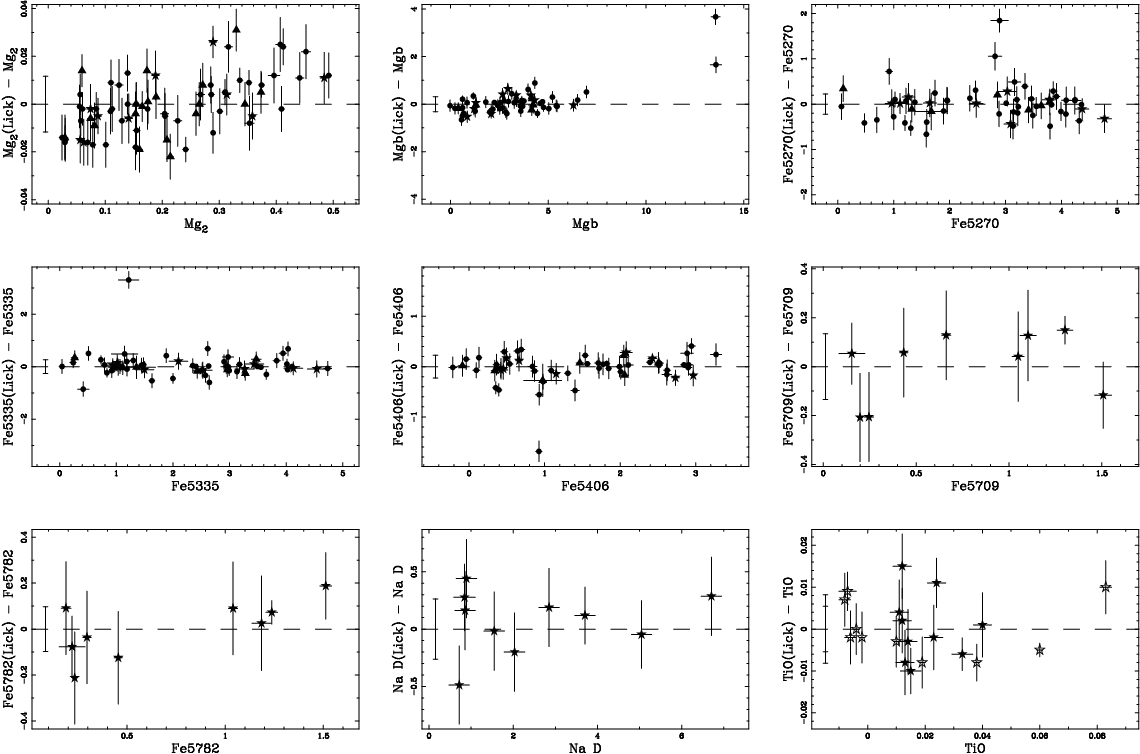}}
 \caption{(Continued)}
\end{figure*}

\subsection{Final data}

Once all the previous corrections were derived, we applied them to the
extracted spectra along the bulge radii and measured the line strengths. We
derived two sets of final data:
\begin{enumerate}
\item In the first one, we introduced the residual errors, masked
the emission lines, converted all the indices to a resolution
corresponding to the maximum velocity dispersion along the radius
(as listed in Table~\ref{tsigmas}), and applied relative offsets
among runs to put them in the same spectrophotometric system
(which should be very close to a true flux-calibrated system). To
do this, we applied the offsets listed in Table 5 with respect to
run 2 (offsets for run 2 were therefore zero). However, we did not
transform the indices to the Lick/IDS system. This data can be
very useful for future studies working at any spectral resolution
and in a flux-calibrated system. Note that, in this dataset,
indices from different galaxies cannot be directly compared, since
indices are measured at different resolutions. \item For the
second dataset, we carried out a full transformation to the
Lick/IDS system, applying (in addition to the corrections from the
last paragraph) the necessary offsets and the index-dependent
velocity dispersion corrections. For this last step, we broadened
the spectra when possible or applied the polynomials of
Table~\ref{sigcoefs} when the original resolutions (the quadratic
sum of $\sigma_{\rm max}$ and the instrumental resolution) were
larger than $\sigma_{\rm L}$ (Table~\ref{indres}).
\end{enumerate}

\subsection{Comparison with other works}
\label{sec.comp}

\begin{figure*}
\resizebox{1.0\hsize}{!}{\includegraphics{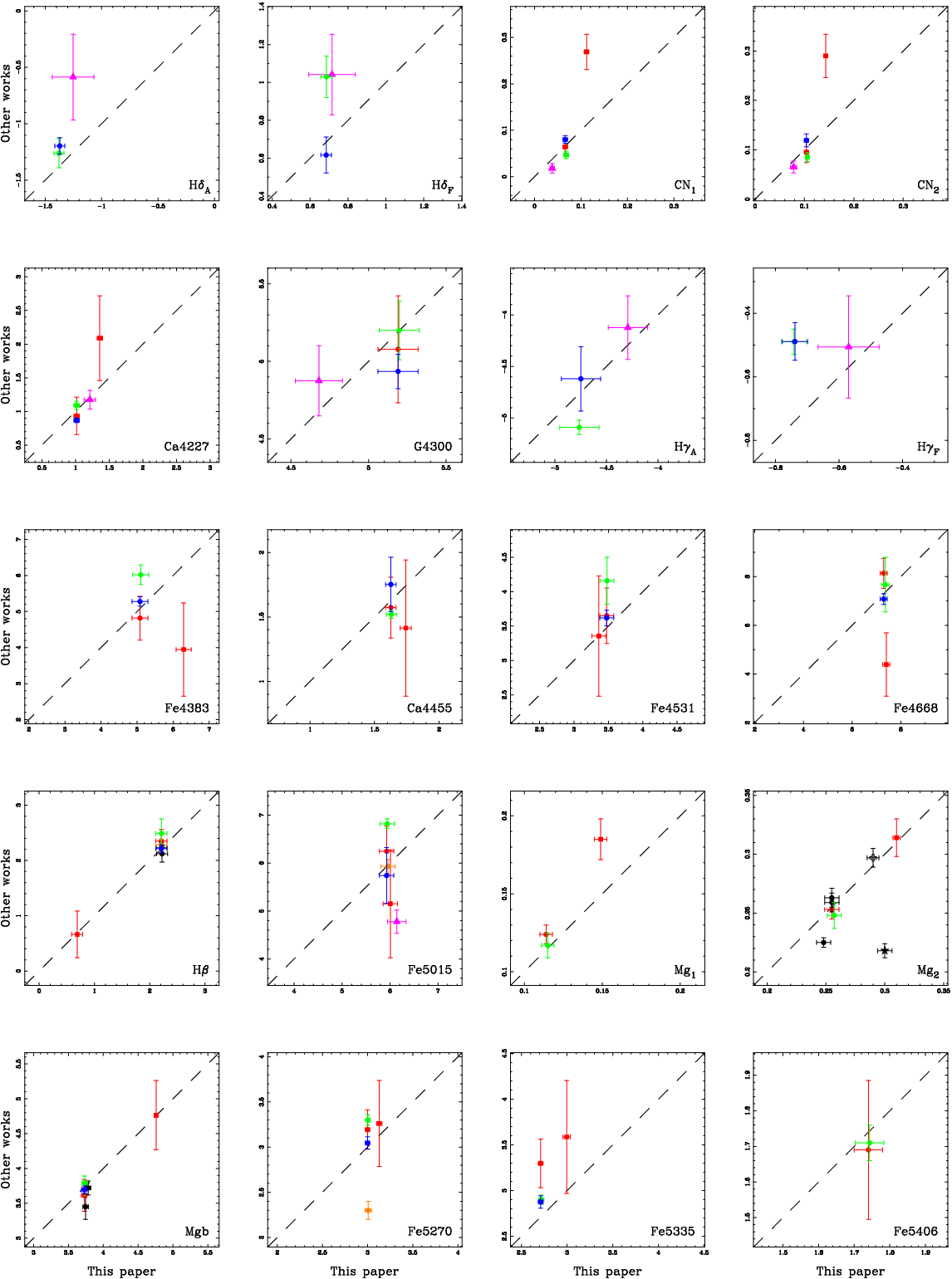}}
 \caption{Comparison of the line-strength indices measured in the
 central regions with data from other workers. Different symbols
 are used for different galaxies: squares for NGC~5084, crosses for NGC~6010,
 triangles for NGC~7703, circles for NGC~7332 and stars for
 NGC~7814. Color codes refer to data from: Trager et al. (1998)
 (red), S\'{a}nchez-Bl\'{a}zquez et al. (2006) (blue), Caldwell et
  al. (2003) (magenta), Kuntschner et al. (2007) (orange),
  Denicol\'{o} et al. (2005) (green), and other sources (black).}
 \label{otherworks}
\end{figure*}

Finally, in Figure~\ref{otherworks} we present a comparison of the
line-strengths measured in the central regions of our bulges with
the available results from the literature. The galaxies with
existing published data and the corresponding references are the
following: NGC~5084 (Trager et al.\ 1998), NGC~6010
(Falc\'{o}n-Barroso et al.\ 2002), NGC~7703 (Caldwell, Rose \&
Concannon 2003), NGC~7332 (Bender, Burstein \& Faber 1993; Trager
et al.\ 1998; Golev et al.\ 1999; Falc\'{o}n-Barroso et al.\ 2002,
2004; Denicol\'{o} et al.\ 2005; Sil'chenko 2006; Kuntschner et
al.\ 2006; S\'{a}nchez-Bl\'{a}zquez et al.\ 2006), and NGC~7814
(Prugniel et al.\ 2001). To perform the comparison, for each
galaxy and reference we binned our spectra in the central regions
of the bulges to reproduce as closely as possible the aperture
used by the corresponding works. It is apparent from the figure
that, although there is a general agreement with the data from
other workers, there are also some important discrepancies for
some particular line-strengths and authors (like some indices for
NGC~5084 from Trager et al.\ 1998, or the Mg$_2$ measurement for
NGC~7814). These differences could be due to: low $S/N$ ratios in
the spectra of other authors, variations in the centering of the
slit or in the observed aperture, or differences in the
calibration and data reduction processes. However, the fact that,
when there are several published values for a galaxy (NGC~7332),
our measurements agree with the majority of them gives us
confidence in our measured values. For instance, for the 5
previous measurements for the central Mg$_2$ index in NGC~7332,
our value agrees within the errors with 4 out of them.

The comparison with the central line-strengths from
S\'{a}nchez-Bl\'{a}zquez et al. (2006) is of particular relevance,
since in Paper~II we will make use of this dataset as a comparison
sample of elliptical galaxies. In this case, in which we plot with
blue symbols the indices measured in an aperture of $2\times4$
arcsec (as listed in S\'{a}nchez-Bl\'{a}zquez 2004), we find a
good agreement with our data (with the exception of the
H$\gamma_{\rm F}$ index, although note that the results agree for
H$\gamma_{\rm A}$).

\section{Summary}  \label{s:summ}
We have introduced our long-term project devoted to understand the evolutionary
status of galaxy bulges by presenting the sample and the most relevant details
of our survey of long-slit spectra along the minor axis of bulges in edge-on
spiral galaxies. After summarizing a data reduction process characterized by a
very detailed error analysis, we have presented our procedures to measure
dynamical parameters and all the Lick/IDS line-strength indices available in
the spectra. A special emphasis has been made in the broadening corrections of
the indices and their conversion to the Lick/IDS system. As a final product, we
have derived tables giving all the parameters (radial velocities, velocity
dispersions and indices) at each galactocentric radius for all the bulges of
the sample. In the case of the indices, two datasets have been obtained: one
with the indices in flux-calibrated spectra and at the {\it natural} resolution
of the bulges (i.e. internal plus instrumental), and another in which all the
indices have been converted to the Lick/IDS system. This two datasets will be
available in electronic form at CDS.

In the next paper of this series we will measure line-strength gradients and
analyze them at the light of different galaxy formation scenarios.

\begin{acknowledgements}
We are grateful to Nicolas Cardiel for his help on technical issues on the
reduction process. The INT is operated on the island of La Palma by the Royal
Greenwich Observatory at the Observatorio del Roque de los Muchachos of the
Instituto de Astrofisica de Canarias. This work was supported by the Spanish
research project AYA 2003-01840.
\end{acknowledgements}

\end{document}